\documentclass[preprint2,twocolumn,times,tighten]{aastex63}

\usepackage{color}
\usepackage{enumitem}
\usepackage{hyperref}
\usepackage{verbatim}
\usepackage{amsmath,amstext,mathrsfs}
\usepackage[all]{hypcap} 
\usepackage{afterpage}    
\usepackage{marginnote}
\usepackage{makecell}
\usepackage{subfigure}
\usepackage{multirow}
\usepackage{lineno}

\shorttitle{Identifying the Gravitational Collapse in Serpens with Velocity Gradients}
\shortauthors{Hu, Lazarian \& Stanimirovi\'{c}}
\graphicspath{{./}{figures/}}

\begin{document}
	
	\title{Revealing Gravitational Collapse in Serpens G3-G6 Molecular Cloud using Velocity Gradients}

	\email{yue.hu@wisc.edu; alazarian@facstaff.wisc.edu}
	
	\author[0000-0002-8455-0805]{Yue Hu}
	\affiliation{Department of Physics, University of Wisconsin-Madison, Madison, WI 53706, USA}
	\affiliation{Department of Astronomy, University of Wisconsin-Madison, Madison, WI 53706, USA}
	\author{A. Lazarian}
	\affiliation{Department of Astronomy, University of Wisconsin-Madison, Madison, WI 53706, USA}
	\affiliation{Center for Computation Astrophysics, Flatiron Institute, 162 5th Ave, New York, NY 10010}
	\author[0000-0002-3418-7817]{Sne\v{z}ana Stanimirovi\'{c}}
	\affiliation{Department of Astronomy, University of Wisconsin-Madison, Madison, WI 53706, USA}

	\begin{abstract}
		The relative role of turbulence, magnetic fields, self-gravity in star formation is a subject of intensive debate.  We present IRAM 30m telescope observations of the $^{13}$CO (1-0) emission in the Serpens G3-G6 molecular cloud and apply to the data a set of statistical methods. Those include the probability density functions (PDFs) of column density and the Velocity Gradients Technique (VGT). We combine our data with the Planck 353 GHz polarized dust emission observations, Hershel H$_2$ column density. We suggest that the Serpens G3-G6 south clump is undergoing a gravitational collapse. Our analysis reveals that the gravitational collapse happens at volume density $n\ge10^3$ $\rm cm^{-3}$. We estimate the plane-of-the-sky magnetic field strength of approximately 120 $\mu G$ using the traditional Davis-Chandrasekhar-Fermi method and 100 $\mu G$ using a new technique proposed in \citet{2020arXiv200207996L}. We find the Serpens G3-G6 south clump's total magnetic field energy significantly surpasses kinetic energy and gravitational energy. We conclude that the gravitational collapse could be successfully triggered in a supersonic and sub-Alfv\'{e}nic cloud.
	\end{abstract}
	

	\section{Introduction}
	\label{sec:intro}
	Star formation in molecular clouds is regulated by a combination of MHD turbulence, magnetic fields, and self-gravity over scales ranging from tens of parsecs to $\ll0.1$ pc \citep{1966ApJ...146..480J,1977ApJ...214..488S,1992pavi.book.....S,1994ApJ...429..781S,1998ARA&A..36..189K,MO07,2011Natur.479..499L,2013ApJ...768..159H,2014ApJ...783...91C,Aetal15,1998ApJ...498..541K}. A number of observational and numerical studies suggest that the interaction of turbulent supersonic flows and magnetic fields produce high-density fluctuations which serve as the nurseries for new stars \citep{1999ApJ...510..274N,2003ApJ...593..426T,MO07,2013MNRAS.436.1245F,2019A&A...623A..16S,Hu20}. However, the balance between the three is still obscured. In particular, it is challenging to identify the gravitational collapse, i.e., where gravity takes over. 
	
	To get insight into self-gravity in molecular clouds, the column density Probability Density Functions (PDFs) are popularly used. In non-gravitational collapsing isothermal supersonic environments, the PDFs appear as log-normal distribution with its width controlled by the sonic Mach number and the driving of the turbulence \citep{1998PhRvE..58.4501P,2000ApJ...535..869K,2008ApJ...680.1083R,2010A&A...512A..81F,2011ApJ...727L..20K,2012ApJ...750...13C,2017ApJ...840...48P,2018ApJ...863..118B}. Self-gravity, further, introduces a power-law tail at high-density range \citep{2011MNRAS.416.1436B,2012ApJ...750...13C,2008ApJ...680.1083R,2018ApJ...863..118B,2019MNRAS.482.5233K}. The transition from log-normal PDFs to power-law PDFs can reveal the density threshold, above which the gas becomes gravitational collapsing. 
	
	The Velocity Gradients Technique (VGT; \citealt{2017ApJ...835...41G,YL17a,LY18a,PCA}, and reference therein) is a novel approach to study the magnetic fields, turbulence, and self-gravity in the interstellar medium (ISM). VGT is rooted in the MHD turbulence theory \citep{GS95} and the turbulent reconnection theory \citep{LV99}. These theories revealed the anisotropic nature of turbulent eddies, i.e., the eddies are elongating along with the local magnetic fields. The magnetic field direction, therefore, is parallel to the eddies' semi-major axis. The gradients of velocity fluctuations, which are perpendicular to the semi-major axis, play the role of probing the magnetic fields \citep{2000ApJ...539..273C,2001ApJ...554.1175M,2002ApJ...564..291C}. Since subsonic velocity and density fluctuations exhibit similar statistical properties, the density gradients (or intensity gradients) are also perpendicular to the magnetic fields \citep{YL17b,IGs}. The above consideration is the basis on which the VGT was developed.
	
	
	However, this perpendicular relative orientation between the gradients and magnetic fields can be broken by self-gravity \footnote{Note that in supersonic turbulence, shocks can also change the relative orientation of density gradients and magnetic fields \citep{YL17b, IGs, 2020MNRAS.498.1593B}.}. In the case of gravitational collapse, the gravitational force pulls the plasma in the direction parallel to the magnetic field producing the most significant acceleration. Because gradients are pointing to maximum changes, the velocity gradients are thus dominated by the gravitational acceleration, being parallel to the magnetic fields \citep{YL17b,Hu20,survey}. For density gradients, the consideration is similar. The most significant accumulation of material happens in the direction of gravitational collapse so that the intensity gradients are parallel to the magnetic fields. This phenomenon has been observed in the molecular clouds Serpens \citep{survey}, G34.43+00.24 \citep{2019ApJ...878...10T}, and NGC 1333 \citep{PDF}. This particular reaction with respect to self-gravity enables VGT to reveal regions of gravitational collapse and quiescent areas where turbulent motions, thermal pressure, and magnetic support dominate over gravitational energy. 
	
	The Serpens cloud, which is famous for its high star formation rate \citep{1999A&A...345..965C} and high surface density of young stellar objects (YSOs; \citealt{2008hsf2.book..693E}), serves as an excellent object to study the gravitational collapse. In particular, the Serpens G3-G6 region \citep{1979ApJS...41..743C} is relatively isolated and does not have nearby \ion{H}{2} regions, or obvious large scale gas shocks, both of which can cause changes in the alignment between magnetic fields, velocity gradients, and density gradients \citep{YL17b, IGs, 2020MNRAS.498.1593B}. Our study, therefore, focuses on the Serpens G3-G6 region. The $\rm ^{12}CO$ (2–1) and $\rm ^{13}CO$ (2–1) emission lines of this region were previously observed with the Arizona Radio Observatory Heinrich Hertz Submillimeter Telescope  \citep{2013ApJS..209...39B} with angular resolution of $\approx38''$. Our new $\rm ^{13}CO$ (1–0) emission line data obtained with the IRAM 30-m telescope complement previous observations and have higher angular resolution ($\approx23.5''$). These data, including the Herschel H$_2$ column density image \citep{2010AA...518L.102A} and the Planck 353 GHz dust polarization data \citep{2020A&A...641A...3P}, were used in this study to resolve kinematic, magnetic, and gravitational properties for Serpens at scales $\le1$ pc. The kinematic property is directly measured from the emission lines' width, while we use two different approaches to estimate either magnetic or gravitational properties. The magnetic field strength was estimated by both the Davis-Chandrasekhar-Fermi method \citep{1951PhRv...81..890D,1953ApJ...118..113C} based on the polarization measurement and a new approach based on the values of sonic Mach number $\rm M_S$ and Alfv\'en Mach number $\rm M_A$ \citep{2020arXiv200207996L}. Also, we employ the PDFs and VGT to confirm the gravitational collapsing nature of the Serpens G3-G6 cloud. 
	
	The paper is organized as follows. In \S~\ref{Sec.data}, we give the details of the observational data used in this work, including data reduction. In \S~\ref{sec:method}, we describe the methodology and algorithms implemented in VGT. In \S~\ref{sec:results}, we present our results of identifying the gravitational collapsing regions on Serpens G3-G6 clump and we analysis the dynamics of the clump. We give discussion in \S~\ref{sec.dis} and conclusion in \S~\ref{sec:con}.
	
	\section{Observations and data reduction}
	\label{Sec.data}
	\subsection{ $\rm ^{13}CO$ (1–0) emission line}
	We obtained a new $\rm ^{13}CO$ (J = 1–0) fully sampled map of Serpens using the IRAM 30m telescope \citep{2012A&A...538A..89C}. The observations were obtained in July 2020 using 16 h of telescope time under average summer weather conditions (6 mm median water vapor). We covered a field of view (FoV) $10'\times40'$. The $\rm ^{13}CO$ (J = 1–0) emission was observed using the EMIR receiver and the VESPA spectrometer using a bandwidth of 60 MHz at 0.092 MHz resolution ($\approx$ 0.212 km s$^{-1}$). The half-power beamwidth (HPBW) at 110.201354 GHz is $\approx 23.5''$. 
	
	We used the on-the-fly scanning strategy with a dump time of
	0.7 s and a scanning speed of $11''$/s to ensure a sampling of three dumps per beam along the scanning direction with the scanning direction reversed after each raster line (i.e., zigzag scanning mode). We covered the full FoV ($\approx$0.11 square degrees) with 15 rectangular tiles (of 15 OTF scans each) along the Declination direction and 20 rectangular tiles (of 5 OTF scans each) along the Right Ascension direction, followed by a calibration measurement. The reference position (REF) was observed for 10 s after each raster line following the pattern
	REF–OTF–OTF–OTF-OTF-OTF-REF along the Declination direction and the pattern REF–OTF–OTF–OTF-REF along the Right Ascension direction. Each tile is of approximately $8''\times40''$ size. In total, we employed about 13 min per tile.
	
	Data reduction was carried out using the GILDAS1/CLASS
	software \footnote{\url{http://www.iram.fr/IRAMFR/GILDAS}}. The data were first calibrated to the $\rm T_A$ scale and were then corrected for atmospheric absorption and spillover losses using the chopper-wheel method  \citep{1973ARA&A..11...51P}. A polynomic baseline of second order was subtracted from each spectrum, avoiding velocities with molecular emission. The spectra were then gridded into a data cube through convolution with a Gaussian kernel of FWHM $\sim 1/3$ of the IRAM-30m telescope beamwidth at the rest line frequency. The typical (1$\sigma$) RMS noise level achieved in the map is 0.33 K per 212 m s$^{-1}$ velocity channel. A large table of the individual spectra was made and the spectra were finally combined to obtain a regularly gridded position-position-velocity data cube, setting the pixel size to 5$''$. Here we convert the measured antenna temperature $\rm T_A$ to brightness temperature $\rm T_b$ through $\rm T_b=(F_{eff}/B_{eff})T_A$, where $\rm F_{eff}=0.95$ is the forward efficiency of the IRAM 30 m telescope and $\rm B_{eff}=0.79$ is the main beam efficiency at 110.201354 GHz \citep{2017A&A...599A..98P}. 
	
	\subsection{ $\rm ^{12}CO$ (2–1) and $\rm ^{13}CO$ (2–1) emission lines}
	The $\rm ^{12}CO$ (2–1) and $\rm ^{13}CO$ (2–1) emission lines were observed with the Heinrich Hertz Submillimeter Telescope \citep{2013ApJS..209...39B} while the H$_2$ column density data is obtained from the Herschel Gould Belt Survey \citep{2010AA...518L.102A}. Each line was measured with 256 filters of 0.25 MHz bandwidth, giving a total spectral coverage of 41 km s$^{-1}$ at a resolution of 0.33 km s$^{-1}$. 
	
	The angular resolution of the emission lines is $38''$ (0.04 pc) with a sensitivity of 0.12 K RMS noise per pixel in one spectral channel \citep{2013ApJS..209...39B}. The radial velocity of the bulk of the emission ranges from about -1 to +18 km s$^{-1}$ for $\rm ^{12}CO$ (2-1) and from +2 to +13 km s$^{-1}$ for $\rm ^{13}CO$ (2-1) \citep{2013ApJS..209...39B}. We select the emissions within these ranges for our analysis.
	
	\subsection{Polarized dust emission}
	To trace the magnetic field orientation in the POS, we use the Planck 353 GHz polarized dust signal data from the Planck 3rd Public Data Release (DR3) 2018 of High Frequency Instrument \citep{2020A&A...641A...3P}\footnote{Based on observations obtained with Planck (\url{http://www.esa.int/Planck}), an ESA science mission with instruments and contributions directly funded by
		ESA Member States, NASA, and Canada.}. 
	
	The Planck observations defines the polarization angle $\phi$ and polarization fraction $p$ through Stokes parameter maps I, Q, and U:
	\begin{equation}
		\begin{aligned}
			\phi&=\frac{1}{2}\arctan(-U,Q)\\
			p&=\sqrt{Q^2+U^2}/I
		\end{aligned}
	\end{equation}
	where $-U$ converts the angle from HEALPix convention to IAU convention and the two-argument function $\arctan$ is used to account for the $\pi$ periodicity. To increase the signal-to-noise ratio, we smooth all maps from nominal angular resolution $5'$ up to a resolution of $10'$ using a Gaussian kernel. The magnetic field angle is inferred from $\phi_B=\phi+\pi/2$.

	\section{Methodology: The Velocity Gradients Technique}
	\label{sec:method}
	
	\subsection{Theoretical consideration}
	The Velocity Gradients Technique (VGT; \citealt{2017ApJ...835...41G,YL17a,LY18a,PCA}) is the main analysis tool in the work. It is theoretically rooted in the advanced magnetohydrodynamic (MHD) turbulence theory (\citealt{GS95}, henceforth GS95), including the concept of fast turbulent reconnection theory (\citealt{LV99}, henceforth LV99). These theories explained that turbulent eddies are anisotropic (see GS95) so that their semi-major-axis is elongating along the local magnetic fields (see LV99). This was numerically demonstrated, by \citet{2000ApJ...539..273C} and \citet{2001ApJ...554.1175M}. In particular, LV99 derived the anisotropy relation for the eddies in local reference frame:
	\begin{equation}
		\label{eq.lv99}
		l_\parallel\simeq L_{inj}(\frac{l_\bot}{L_{inj}})^{\frac{2}{3}}{\rm M_A^{-4/3}}
	\end{equation}
	where $l_\bot$ and $l_\|$ are the perpendicular and parallel size of eddies with respect to the {\it local} magnetic field. $\rm M_A$ is the Alfv\'{e}n Mach number and $L_{inj}$ is the injection scale of turbulence. The corresponding scaling of velocity fluctuation $v_l$ at scale $l$ is then:
	\begin{equation}
		\label{eq.v}
		v_l\simeq v_{inj}(\frac{l_\bot}{L_{inj}})^{\frac{1}{3}}{\rm M_A^{1/3}}
	\end{equation}
	where $v_{inj}$ is the injection velocity. Explicitly, as the anisotropic relation indicates $l_\bot \ll l_\parallel$, the velocity gradients scale as \citep{YL20a}:
	\begin{equation}
		\label{eq.grad}
		\begin{aligned}
			\nabla v_l&\propto\frac{v_{l}}{l_\bot}\simeq \frac{v_{inj}}{L_{inj}}(\frac{l_{\perp}}{L_{inj}})^{-\frac{2}{3}}{\rm M_A^{\frac{1}{3}}}
		\end{aligned}
	\end{equation}
	which means that the smallest resolved eddies induce
	the largest gradients. The gradients are perpendicular to the
	local magnetic fields.
	
	\subsection{Principal component analysis}
	As we discussed above, the velocity fluctuation, i.e., the velocity variance, is related to the eddy's size along the LOS. Large velocity variance eddies give the most significant contribution to the velocity gradients. To get the largest velocity variance, \citet{PCA} proposed the Principal Component Analysis (PCA) to pre-process the spectroscopic PPV cubes. 
	
	PCA treats the observed brightness temperature $T_b(x, y, v)$ in PPV cube as the probability density function of three random variables $x$, $y$, $v$. By splitting the PPV cube into $n_v$ velocity channels along the LOS, the covariance matrix $S(v_i,v_j)$ and its corresponding eigenvalue equation are defined as:
	\begin{equation}
		\label{eq:13}
		\begin{aligned}
			S(v_i,v_j) \propto &\int dxdy T_b(x,y,v_i)T_b(x,y,v_j)\\
			&-\int dxdy T_b(x,y,v_i)\int dxdyT_b(x,y,v_j)
		\end{aligned}
	\end{equation}
	\begin{equation}
		\label{eq:14}
		\textbf{S}\cdot \textbf{u}=\lambda\textbf{u}
	\end{equation}
	here $i,j=1,2,...,n_v$ and $\lambda$ is the eigenvalues associated with the eigenvector $\textbf{u}$. The equation gives totally $n_v$ eigenvalues. Each eigenvalue $\lambda_i$ is proportional to the squared velocity variance $v_i^2$ along the LOS. The velocity channel corresponding to the largest eigenvalue, therefore, has the most significant contribution to the velocity gradients. 
	
	In the frame of VGT, the PPV cube can be pre-processed by PCA in two different ways. The first one is treating the eigenvalues $\lambda$ as the weighting coefficients for each velocity channel \citep{PCA}. By integrating the weighted velocity channels along the LOS, we obtained the velocity centroid map \textbf{C(x,y)}:
	\begin{equation}
		C(x,y)=\frac{\int dvT_b(x,y,v)\cdot v\cdot\lambda(v)}{\int dvT_b(x,y,v)}
	\end{equation}
	Large eigenvalues enhance the significance of their corresponding velocity channels, while small eigenvalues give a suppression.
	
	Alternatively, instead of weighting velocity channels, the PPV cube can be projected into a new orthogonal basis constructed by $n_v$ eigenvectors \citep{EB}. Its corresponding eigenvalue constrains the length of each eigenvector. Since the new axes are oriented along the direction of maximum variance, the PPV cube on a new basis exhibits the largest velocity variance. The projection of is operated by weighting channel $T_b(x,y,v_j)$ with the corresponding eigenvector element $u_{ij}$, in which the corresponding eigen-channel $I_i(x,y)$ is:
	\begin{equation}
		I_i(x,y)=\sum_j^{n_v}u_{ij}\cdot T_b(x,y,v_j)
	\end{equation}
	This step produces a total of $n_v$ eigen-channels in the eigenvectors space. Generally, both the weighting and projecting approaches are figuring out the maximum velocity variance, which is the most important in velocity gradients' calculation. In this work, we adopt the projection approach to construct the pseudo-Stokes-parameters; see the discussion below.
	
	Note here the channel number $n_v$ should be sufficiently large so that the channel width $\Delta v$ satisfies $\Delta v<\sqrt{\delta (v^2)}$, where $\sqrt{\delta (v^2)}$ is the 1D velocity dispersion. We call the channel, which satisfies the criteria as a thin velocity channel. Due to the velocity caustic effect, which comes from the non-linear mapping from real space to PPV space, the thin velocity channel's intensity fluctuation is mainly induced by velocity fluctuation instead of density fluctuation \citep{LP00}. The thin velocity channels, therefore, record velocity information, which is used to calculate velocity gradients.
	
	\subsection{Sub-block averaging}
	The PPV cubes are pre-processed by PCA to extract the most crucial velocity components resulting in $n_v$ eigen-channels. Each eigen-channel is convolved with 3 $\times$ 3 Sobel kernels\footnote{$$
		G_x=\begin{pmatrix} 
			-1 & 0 & +1 \\
			-2 & 0 & +2 \\
			-1 & 0 & +1
		\end{pmatrix} \quad,\quad
		G_y=\begin{pmatrix} 
			-1 & -2 & -1 \\
			0 & 0 & 0 \\
			+1 & +2 & +1
		\end{pmatrix}
		$$}
	$G_x$ and $G_y$ to calculate pixelized gradient map $\psi_{g}^i(x,y)$:
	\begin{equation}
		\label{eq:conv}
		\begin{aligned}
			\bigtriangledown_x I_i(x,y)&=G_x * I_i(x,y)  \\  
			\bigtriangledown_y I_i(x,y)&=G_y * I_i(x,y)  \\
			\psi_{g}^i(x,y)&=\tan^{-1}\left(\frac{\bigtriangledown_y I_i(x,y)}{\bigtriangledown_x I_i(x,y)}\right) \, ,
		\end{aligned}
	\end{equation}
	where $\bigtriangledown_x I_i(x,y)$ and $\bigtriangledown_y I_i(x,y)$ are the $x$ and $y$ components of gradient respectively, and $*$ denotes the convolution.

	Note that the turbulent eddy discussed above is a statistical concept. A single gradient, therefore, does not necessarily correlate to the magnetic field. The orthogonal relative orientation between velocity gradients and the magnetic field appears only when the sampling is statistically sufficient. The statistical sampling procedure is proposed by \citet{YL17a}, i.e., the sub-block averaging method. Firstly, the sub-block averaging method takes all gradients orientation within a sub-block of interest and then plots the corresponding histogram. A Gaussian fitting is then applied to the histogram. The Gaussian distribution's expectation value is the statistically most probable gradient's orientation within that sub-block. Incidentally, the sub-block averaging also partially suppresses the RMS noise in the spectroscopic data.
	
	The selection of sub-block size, which controls the number of sampling points, is crucial. \citet{Hu20} later improved the sub-block averaging method to be adaptive. The sub-block centers were selected continuously, locating at the position (i, j), i, j = 1, 2, 3, etc. All gradients pixels within the rectangular boundaries [i-d/2, i+d/2] and [j-d/2, j+d/2] are selected to do averaging, where d is the sub-block size.  We vary the sub-block size and check its corresponding fitting errors within the 95\% confidence level. When the fitting error reaches its minimum value, the corresponding sub-block size is the optimal
	selection. We refer to this procedure as the adaptive sub-block (ASB) averaging method \citep{Hu20}.
	
	\subsection{Pseudo-Stokes-parameters}
	By repeating the gradient's calculation and ASB for each eigen-channel, we obtain totally $n_v$ eigen-gradient maps $\psi_{gs}^i(x,y)$ with $i=1,2,...,n_v$. In analogy to the Stokes parameters of polarization, the pseudo Q$_g$ and U$_g$ of gradient-inferred magnetic fields are defined as:
	\begin{equation}
		\label{eq.qu}
		\begin{aligned}
			& Q_g(x,y)=\sum_{i=1}^{nv} I_i(x,y)\cos(2\psi_{gs}^i(x,y))\\
			& U_g(x,y)=\sum_{i=1}^{nv} I_i(x,y)\sin(2\psi_{gs}^i(x,y))\\
			& \psi_g=\frac{1}{2}\tan^{-1}(\frac{U_g}{Q_g})
		\end{aligned}
	\end{equation}
	The pseudo polarization angle $\psi_g$ is then defined correspondingly. Similar to the Planck polarization, $\psi_B=\psi_g+\pi/2$ gives the POS magnetic field orientation.
	
	\subsection{Alignment measure}
	
	The relative alignment between magnetic fields orientation inferred from Planck polarization $\phi_B$ and velocity gradient $\psi_B$ is quantified by the \textbf{Alignment Measure} (AM, \citealt{2017ApJ...835...41G}): 
	\begin{align}
		\label{AM_measure}
		AM=2(\langle \cos^{2} \theta_{r}\rangle-\frac{1}{2})
	\end{align}
	where $\theta_r=|\phi_B-\psi_B|$ and$\langle ...\rangle$ denotes the average within a region of interests. The value of AM spans from -1 to 1.
	AM = 1 mean $\phi_B$ and $\psi_B$ are parallel, while AM = -1 indicates $\phi_B$ and $\psi_B$ are perpendicular. The standard deviation divided by the sample size's square root gives the uncertainty $\sigma_{AM}$.
	
	\subsection{Double-peak histogram}
	The double-peak histogram achieves the detection of the gravitational collapsing region.  Unlike the gradients $\psi_B$ in diffuse regions, the gravitational collapse flips the gradients by 90$^\circ$ being $\psi_B+\pi/2$. The histogram of gradients' angle can be used to extract this change. In the diffuse region, the histogram exhibits a single Gaussian peak locates at $\psi_B$. The single peak of the histogram becomes $\psi_B+\pi/2$ in gravitational collapsing regions. However, in transitional regions, which cover, for instance, half diffuse material and half gravitational collapsing, the histogram is therefore expected to show two peak values of $\psi_B$ and $\psi_B+\pi/2$. We call this type of histogram the double-peak histogram (DPH).
	
	The DPH works as follows. Every single pixel of $\psi_B$ is defined as the center of a sub-block\footnote{Note this sub-block is different from the ASB, which is used to determine the mean gradients' orientation. The second sub-block implemented in the DPH is used to extract the change of gradients. The second sub-block size can be different from the one for the ASB. In this work, we adopt the second block size $30\times30$ pixels, which are statistically sufficient \citet{Hu20}.} to draw the histogram. An envelope, which is a smooth curve outlining the extremes, is adopted for the histogram to suppress noise and the effect from insufficient bins. Any term whose histogram weight is less than the mean weight value of the envelope is masked. After masking, we work out the peak value of each consecutive envelope. Once more than one peak values appear, and the maximum difference is within the range 90$^\circ\pm\sigma_g$, where $\sigma_g$ is the total standard deviation, the center of this second sub-block is labeled as the boundary of a gravitational collapsing region. The details of the algorithm are presented in \citet{Hu20}. 
	\begin{figure}
		\centering
		\includegraphics[width=1.0\linewidth,height=0.61\linewidth]{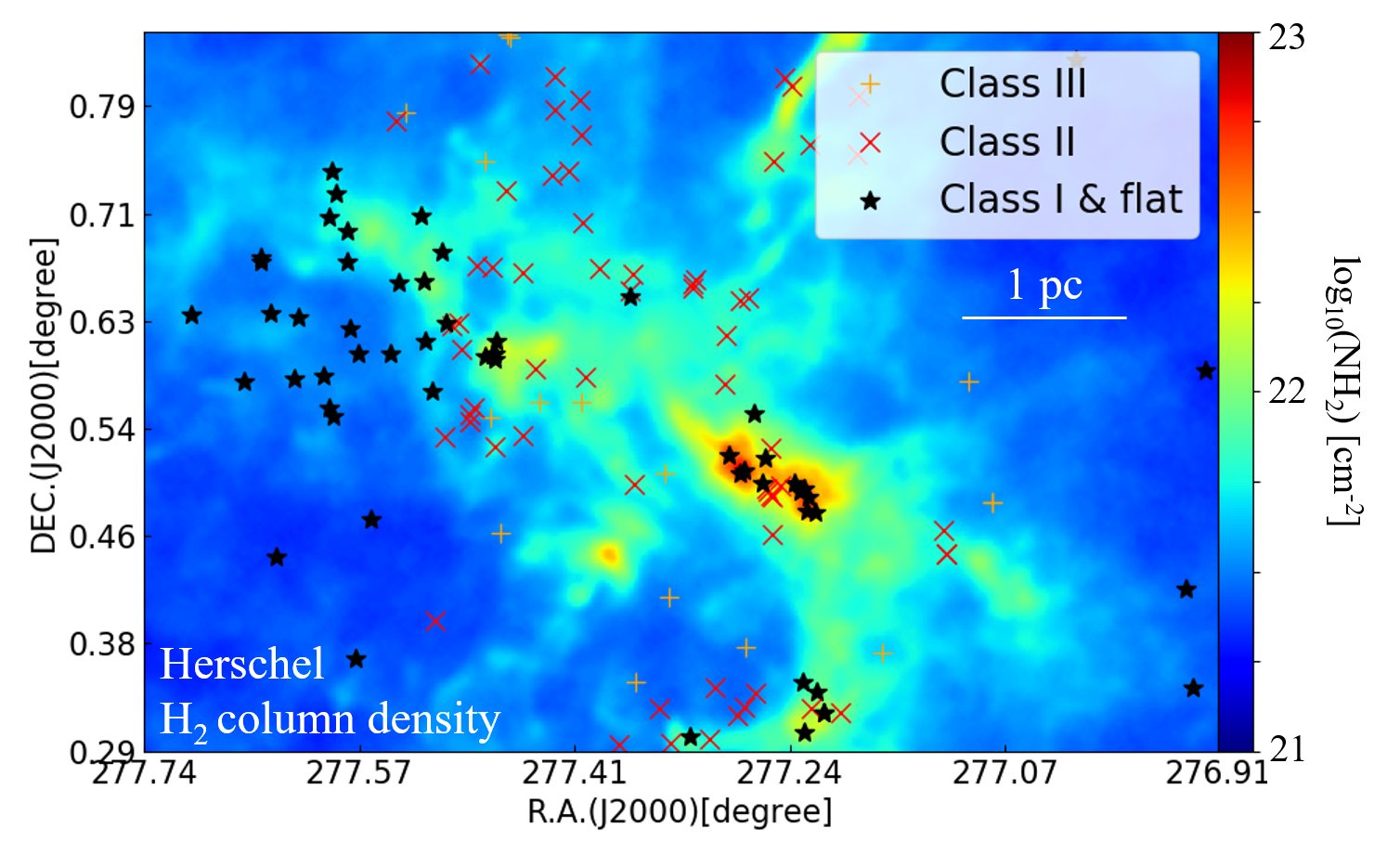}
		\caption{\label{fig:her} Map of $\rm H_2$ column density obtained from the Herschel Gould Belt Survey. The YSOs are identified by \citet{2007ApJ...663.1149H}.}
	\end{figure}
	
	\begin{figure*}
		\centering
		\includegraphics[width=1.0\linewidth,height=0.5\linewidth]{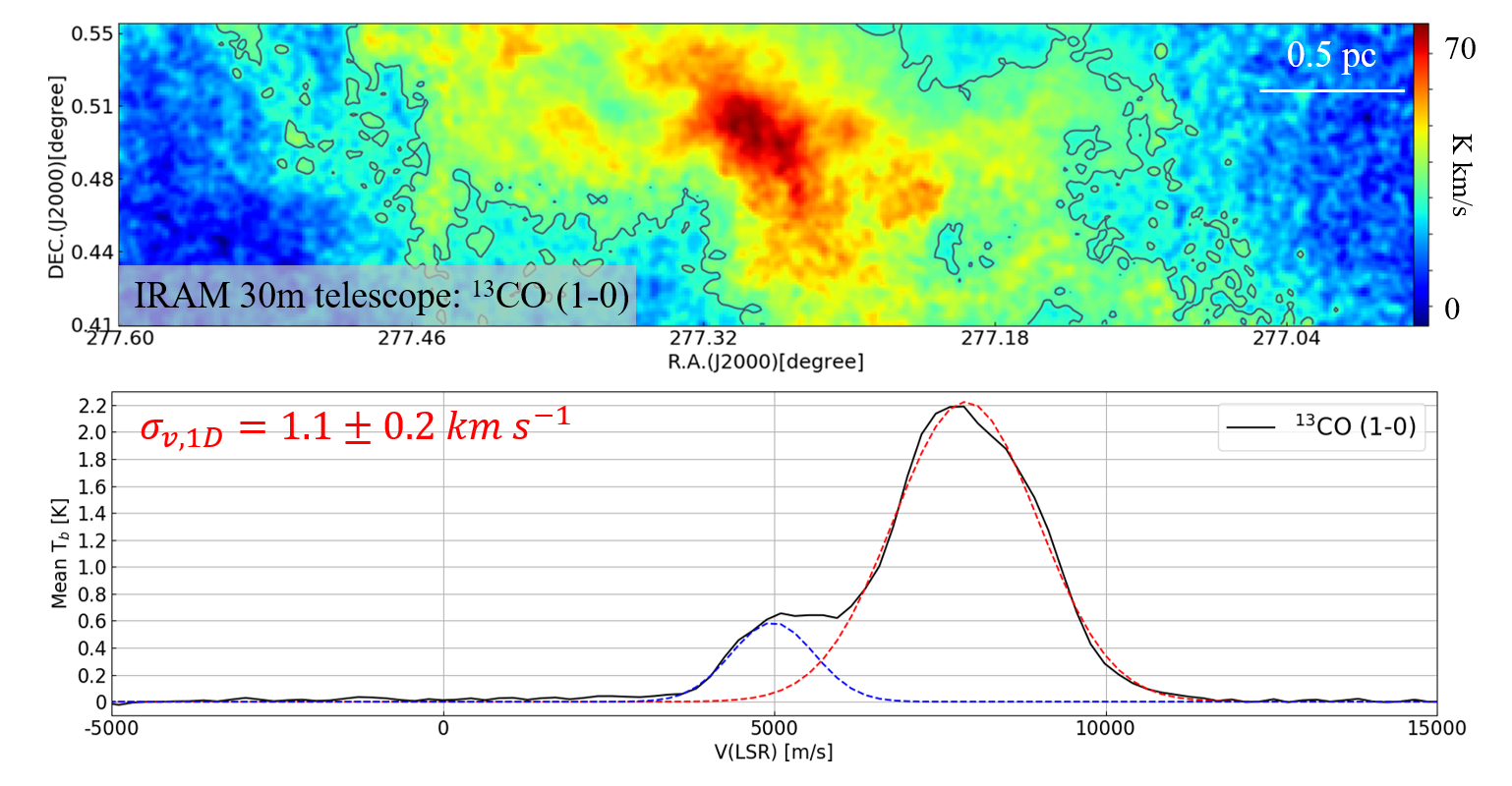}
		\caption{\label{fig:CO_map} Map of integrated brightness temperature of $\rm ^{13}CO$ J = 1–0 (top) over velocity range 2.6 to 12.0 km s$^{-1}$ and the mean brightness temperature spectra (bottom) averaged over the region. The emission line is observed with the IRAM 30m telescope. The contour indicates a mean intensity value of 28.68 K km/s.
		}
	\end{figure*}
	\begin{figure*}
		\centering
		\includegraphics[width=1.0\linewidth,height=0.23\linewidth]{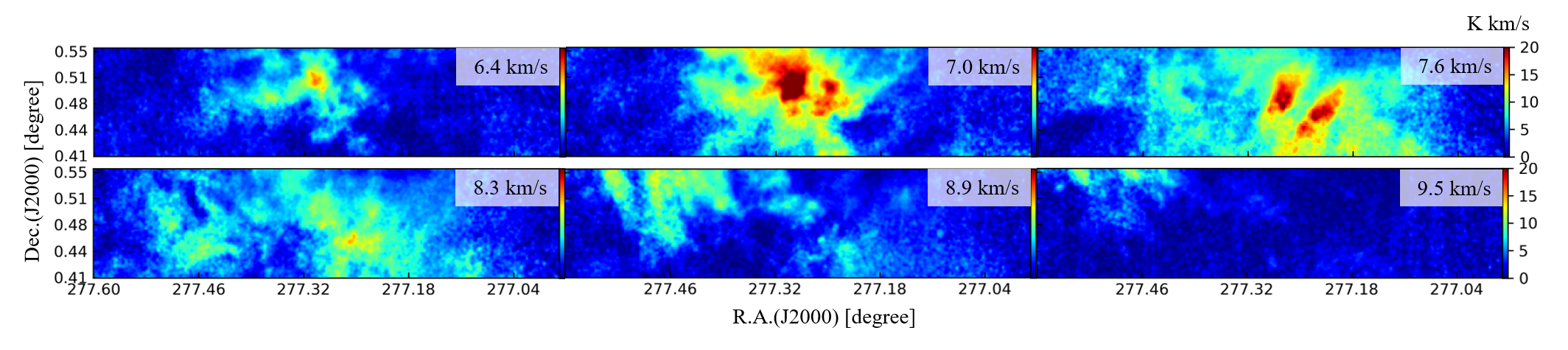}
		\caption{\label{fig:channel} $\rm ^{13}CO$ J = 1–0 spectral channel maps averaged over 636 m s$^{-1}$ and spaced 212 m s$^{-1}$ apart. Mean LSR velocity is labeled in upper right.}
	\end{figure*}
	\section{Results}
	\label{sec:results}
	\subsection{Integrated intensity map and H$_2$ column density map}
	Fig.~\ref{fig:her} presents H$_2$ column density map for the Serpens G3-G6 clump. The H$_2$ column density structures are elongated and filamentary. Here we plot the distribution of young stellar objects (YSOs) within this clump. The YSOs are identified by \citet{2007ApJ...663.1149H}. The evolutionary stage of a YSO can be classified as (in order of youngest to oldest): Class I, flat, Class II, and Class III \citep{1987IAUS..115....1L,1994ApJ...420..837A,1994ApJ...434..614G}. We find Class I and flat YSOs concentrate on the south dense clump core and the north-east filamentary tail. Class II YSOs' mainly locate at the north filamentary tail, while Class III only occupies a small fraction. YSOs' high surface density at the dense clump core and the north tail indicates these two regions are actively forming stars.
	
	The $\rm ^{13}CO$ (1-0) emission line observed with the IRAM 30-m telescope zooms into the south dense clump with a resolution of $23.5''$ ($\approx0.025$ pc, see Fig.~\ref{fig:CO_map}). The radial velocity of the emission ranges from about 2.6 to 12.0 km s$^{-1}$. The spectral line appears to have two apparent peaks : 2.22 K at 7.65 km s$^{-1}$ and 0.58 K at 4.9 km s$^{-1}$. After fitting a double Gaussian profile to the integrated spectral line shown in Figure 2, we obtain the 1D velocity dispersion $\sigma_{v,1D}=1.1\pm0.2$ km s$^{-1}$ for the dominating emission feature. Fig.~\ref{fig:CO_map} also shows the integrated intensity maps for the south dense clump core of Serpens G3-G6. The integration of $\rm ^{13}CO$ (1-0) considers pixels where the brightness temperature is larger than 0.9 K, which is about three times the RMS noise level.  The $\rm ^{13}CO$ (1-0) emission lines cover a wider $ 0.60^\circ\times0.15^\circ$ area. The clump's $\rm ^{13}CO$ (1-0) is still filamentary, spanning from north to south.
	\begin{figure}
		\centering
		\includegraphics[width=1.0\linewidth,height=1.86\linewidth]{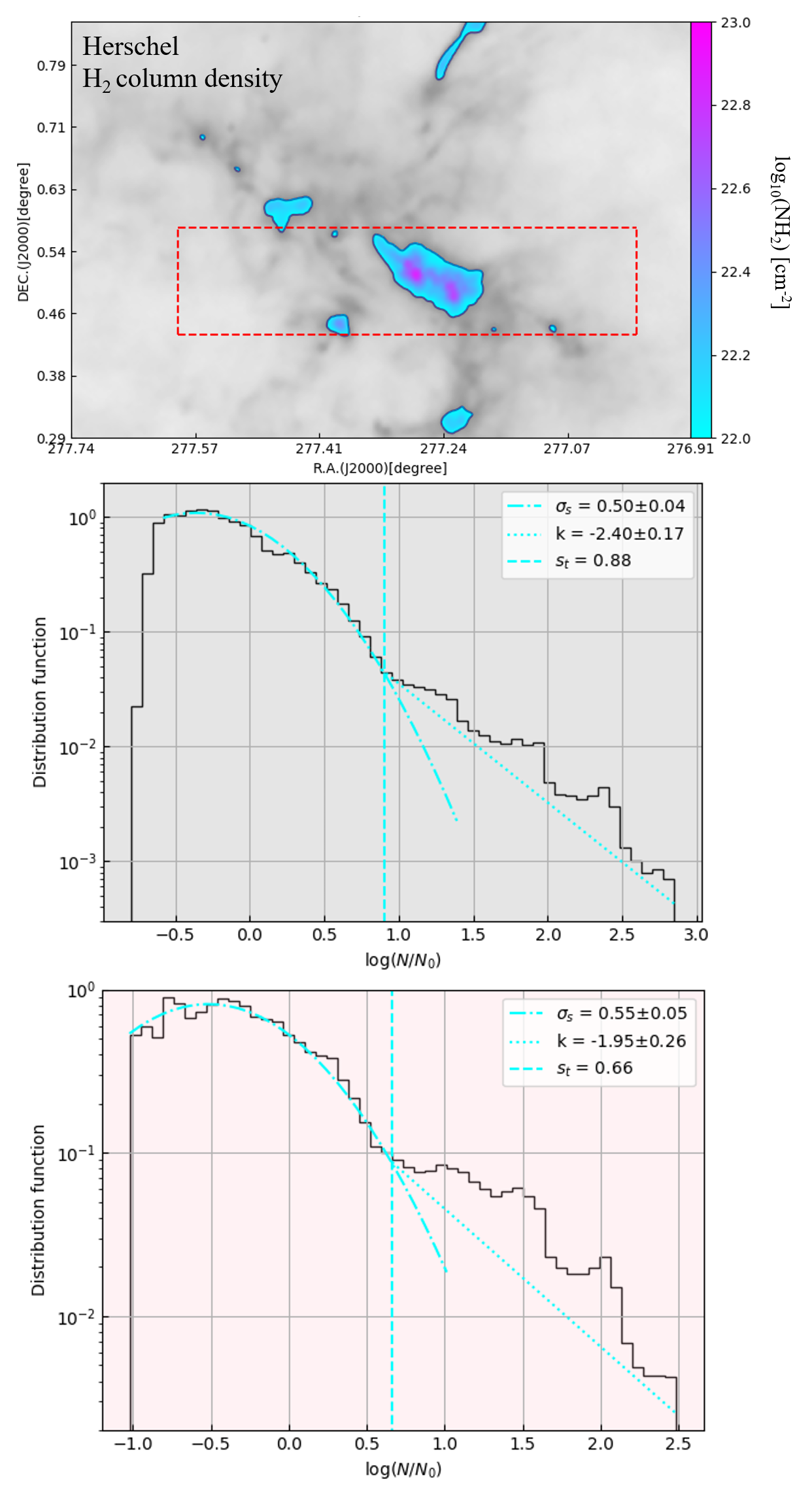}
		\caption{\label{fig:NPDF}\textbf{Top:} Map of of $\rm H_2$ column density. Cyan areas indicate the gravitational collapsing regions identified from the PDFs. \textbf{Middle:} The PDFs of entire column density map. \textbf{Bottom:} The PDFs of south dense clump (see the red box on the top map). $\sigma_s$ gives the dispersion of the log-normal distribution. $S_t$ is the transition density threshold and $k$ is the slope of the reference power-law line.}
	\end{figure}
	
	Fig.~\ref{fig:channel} shows representative velocity channels from the $\rm ^{13}CO$ (1-0) emission line data, averaged over 636 m s$^{-1}$. The intensity structures seen in velocity channels do not appear filamentary. In particular, two distinct dense structures appear at 7.0 km s$^{-1}$ and 7.6 km s$^{-1}$. These structures suggest that the velocity caustic effect is significant, i.e., velocity fluctuations dominate the thin velocity channels \citep{LP00}. 
	
	\subsection{gravitational collapsing regions identified from PDFs}
	\label{subsec:pdf}
	The column density PDFs are widely used to study turbulence and self-gravity in ISM. The PDFs in gravitational collapsing isothermal supersonic turbulence follow a hybrid of log-normal distribution $P_{LN}(s)$ in low-intensity range and power-law distribution $P_{PL}(s)$ in high-intensity range \citep{2008ApJ...680.1083R,2011MNRAS.416.1436B,2012ApJ...750...13C,2018ApJ...863..118B,2019MNRAS.482.5233K}:
	\begin{equation}
		\begin{aligned}
			P_{LN}(s)&=\frac{D}{\sqrt{2\pi\sigma_s^2}}e^{-\frac{(s-s_0)^2}{2\sigma_s^2}},s<S_t\\
			P_{PL}(s)&=DC e^{k s},s>S_t\\
		\end{aligned}
	\end{equation}
	where s$ = ln(N/N_0)$ is the logarithm of the normalized column density $N$. S$_t$ is transitional density between the $P_{LN}$ and $P_{LN}$. s$_0=-\frac{1}{2}\sigma_s^2$ is the mean logarithmic density. The standard deviation $\sigma_s$ for magnetized turbulence is related to the sonic Mach number $\rm M_S$, driving parameter $b$, and compressibility $\beta$ \citep{2012arXiv1203.2117M}:
	\begin{equation}
		\label{eq.b}
		\sigma_s^2=\log(1+b^2{\rm M_S}^2\frac{\beta}{\beta+1})    
	\end{equation}
	The turbulence driving parameter $b=1/3$ for purely solenoidal driving, while $b=1$ for purely compressive driving. For a natural mixture of solenoidal and compressive driving, b is $\approx0.4$ \citep{2015MNRAS.448.3297F,2016ApJ...832..143F}. Assuming $P_{NL}(s)+P_{PL}(s)$ is normalized, continuous, and differentiable, the coefficient D and C are \citep{2017ApJ...834L...1B,2018ApJ...863..118B}:
	\begin{equation}
		\begin{aligned}
			D&=(\frac{Ce^{S_tk}}{-k}+\frac{1}{2}+\frac{1}{2}{\rm erf}(\frac{2S_t+\sigma_s^2}{s\sqrt{2}\sigma_s}))\\
			C&=\frac{e^{0.5(k+1)k\sigma_s^2}}{\sigma_s\sqrt{2\pi}}
		\end{aligned}
	\end{equation}
	
	In Fig.~\ref{fig:NPDF}, we plot the H$_2$ column density PDFs for the entire Serpens G3-G6 clump. The PDFs are log-normal with dispersion $\sigma_s\approx0.50\pm0.04$ till $S_t\approx0.88$, which indicates the gas is gravitational collapsing when its density is larger than $\rm N=e^{S_t}N_0\approx1.02\times10^{22}$ cm$^{-2}$. This column density threshold reveals that the south dense clump is gravitational collapsing.
	The characteristic slope $k$ of that power-law is -2.40$\pm0.17$. Also, we zoom into the south dense region (see the red box in Fig.~\ref{fig:NPDF}), which corresponding to the measured $\rm ^{13}CO$ (1-0). The south dense clump's PDFs are still the combination of $P_{LN}$ and $P_{PL}$. The transition density $S_t\approx0.66$ and the characteristic slope $k=-1.95\pm0.26$ identified the same gravitational collapsing regions as before. The characteristic slope is related to the cloud mean free-fall time, the magnetic fields, and the efficiency of feedback \citep{2013ApJ...763...51F,2014ApJ...781...91G,2017ApJ...834L...1B,2018MNRAS.477.5139G}. For instance, $k=-2.40\pm0.17$ and $k=-1.95\pm0.26$ correspond to  radial density distributions $\rho(r)\propto r^{-1.83\pm0.06}$ and $\rho(r)\propto r^{-2.03\pm0.14}$, respectively\footnote{Here we use the relation $\rho(r)\propto r^{-\chi}$ and $P_{PL}(s)\propto e^{ks}= e^{\frac{2}{1-\chi}s}$ (see Eq.11 in \citealt{2013ApJ...763...51F}).}. The
	power-law slopes of radial density distributions suggest that all the high density gas having collapsed into isothermal cores $\rho(r)\propto r^{-2}$ \citep{1977ApJ...214..488S} and star formation efficiency is in the range of 0\%–20\% \citep{2013ApJ...763...51F}.
	\begin{figure}
		\centering
		\includegraphics[width=1.0\linewidth,height=1.2\linewidth]{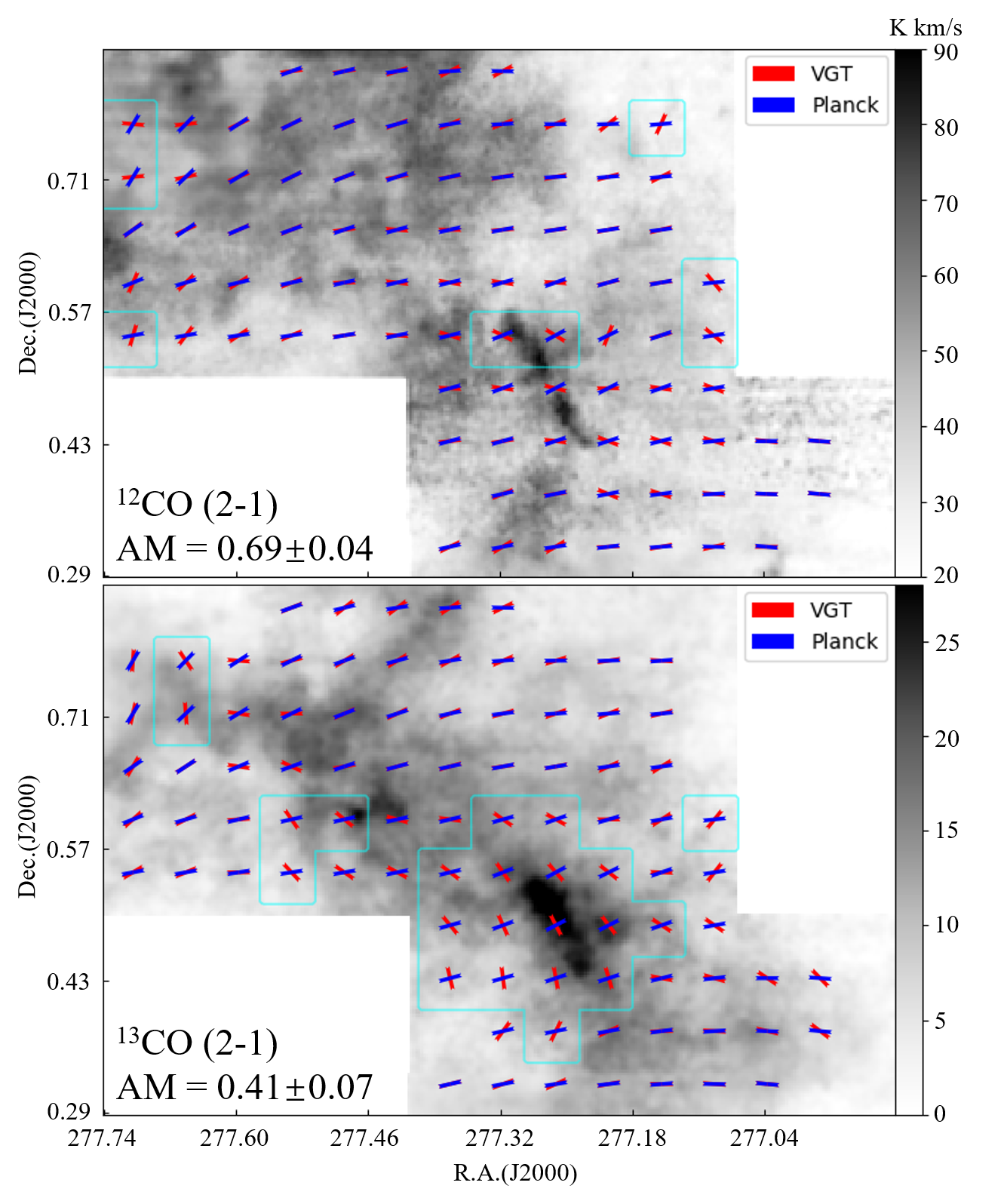}
		\caption{\label{fig:HHT_vgt} \textbf{Top:} The POS magnetic fields inferred from VGT (red segments) using $\rm ^{12}CO$ (2-1) emission line and Planck 353 GHz polarization (blue segments). \textbf{Bottom:} The POS magnetic fields inferred from VGT (red segments) using $\rm ^{13}CO$ (2-1) emission line and Planck 353 GHz polarization (blue segments). Cyan outlines indicate AM $<$ 0 areas. }
	\end{figure}
	
	\begin{figure}
		\centering
		\includegraphics[width=1.0\linewidth,height=1.44\linewidth]{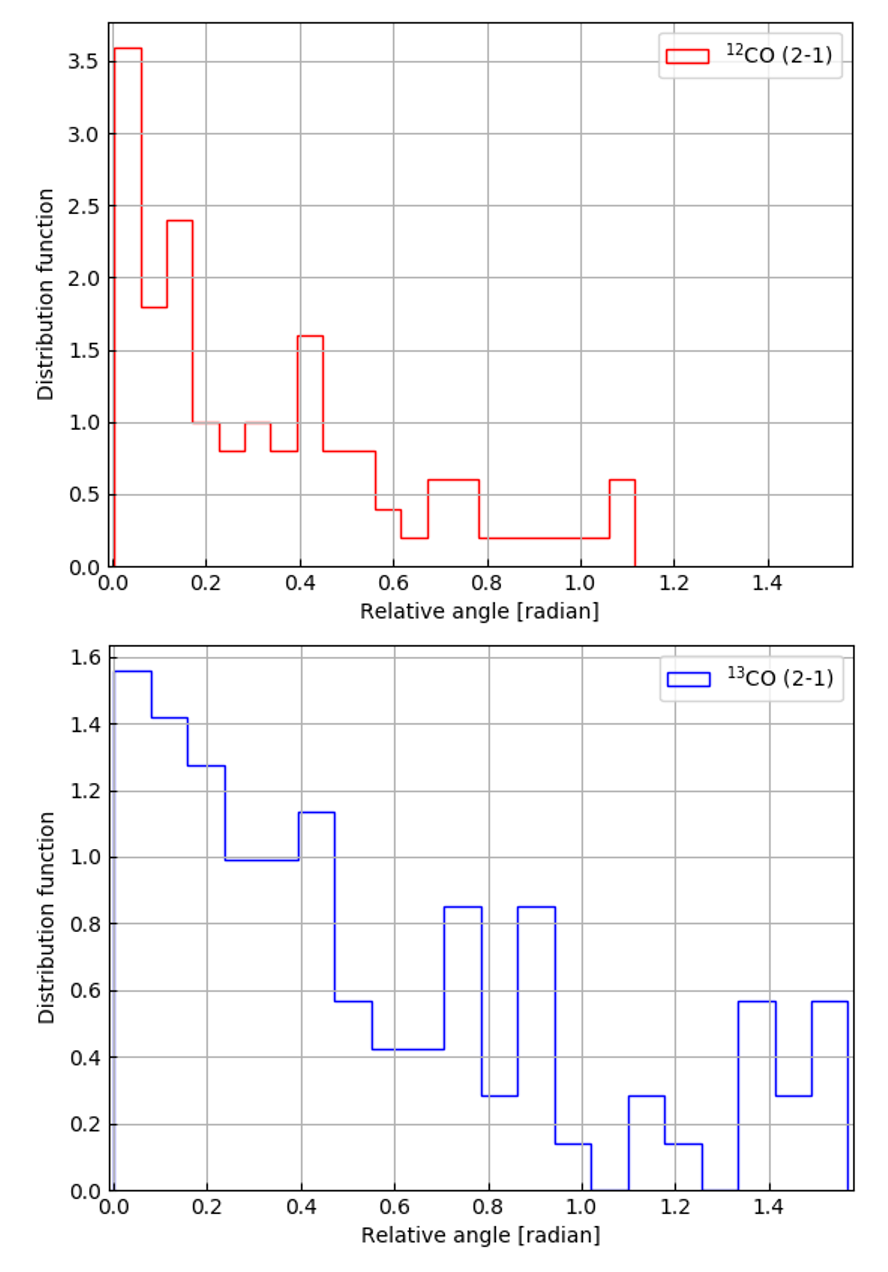}
		\caption{\label{fig:HRO} The histograms of relative angle between the magnetic field inferred from Planck polarization and VGT using $\rm ^{12}CO$ (top) and $\rm ^{13}CO$ (bottom).}
	\end{figure}
	
	
	\subsection{Magnetic fields morphology traced by VGT}
	\label{subsec:B_vgt}
	In Fig.~\ref{fig:HHT_vgt}, we present the results of the VGT analysis using the $\rm ^{12}CO$ (2-1) and $\rm ^{13}CO$ (2-1) emission lines data sets. We use the emission within the velocity ranges [+1, +13] km s$^{-1}$ for calculation following the recipe presented in Sec.~\ref{sec:method}. Pixels, where brightness temperature is less than three times the  RMS noise level (0.4 K), are blanked out.  We average the gradients over $20\times20$ pixels and smooth the gradients map $\psi_g$ (see Eq.\ref{eq.qu}) with a Gaussian filter with kernel width 1, i.e., over 5 pixels.
	
	We also compare the magnetic field inferred from the Planck 353 GHz polarized dust signal data (FWHM $\approx10'$). We re-grid the Planck polarization further to achieve the same pixel size as emission lines. The polarization vector is also averaged over $20\times20$ pixels to match the gradients map. We find the resulting gradients of $\rm ^{12}CO$ (2-1) have good agreement with the Planck polarization, showing AM = 0.69$\pm0.04$. Several anti-alignment vectors appear in the south dense region and image boundary. It is likely that the dust polarization and the $\rm ^{12}CO$ (2-1) emission probe different spatial regions. The optically thick tracer $\rm ^{12}CO$ samples the outskirt diffuse region of the cloud with volume density $n\approx10^{2}$ $\rm cm^{-3}$, while dust polarization likely traces denser regions. The theory of Radiative Torque (RAT) alignment (see \citealt{2007ApJ...669L..77L,Aetal15} for a review) predicts that dust grains can remain aligned at high densities, especially in the presence of the embedded stars. This is in agreement with numerical simulations  \citep{2007ApJ...663.1055B,2019MNRAS.482.2697S}. Therefore, we expect that for $n\ge10^{3}$ $\rm cm^{-3}$ grains are aligned in the regions that we study. In addition, the low resolution ($10'$) of the Planck data also contributes to the misalignment. The VGT measurements from $\rm ^{13}CO$ (2-1) gives less alignment (AM = 0.41$\pm0.07$) with the Planck polarization. In particular, the gradients become perpendicular to the magnetic field in the south dense region. As the south dense region is identified to be gravitational collapsing by the PDFs (see \S~\ref{subsec:pdf}), this change the gradients' direction suggests that presence of a gravitational collapse(see \S~\ref{sec:method}). Since the overall magnetic fields horizontally cross the clump, while the gradients are nearly vertical, the resolution effects do not change this conclusion. 
	
	We plot the histograms of the relative alignment between rotated gradients and the magnetic fields in Fig.~\ref{fig:HRO}. The histogram based on the $\rm ^{12}CO$ (2-1) data is more concentrated around 0, which means that two vectors are parallel. The distribution based on the $\rm ^{13}CO$ (2-1) data, however, spreads all the way to $\pi/2$, which means that two vectors are perpendicular. Previous studies reported that the VGT results of the $\rm ^{13}CO$ emission are more accurate than the ones derived using the $\rm ^{12}CO$ emission when comparing with dust polarization \citep{velac,2020arXiv200715344A} in non-gravitational collapsing clouds. Here we find that the VGT of $\rm ^{12}CO$ gives better alignment. It is likely because the gravitational collapse in the Serpens G3-G6 south clump happens in dense gas $n\ge10^{3}$ $\rm cm^{-3}$. The $\rm ^{12}CO$ emission samples more diffuse regions so that its gradients are less affected by self-gravity.
	\begin{figure*}
		\centering
		\includegraphics[width=0.80\linewidth,height=0.52\linewidth]{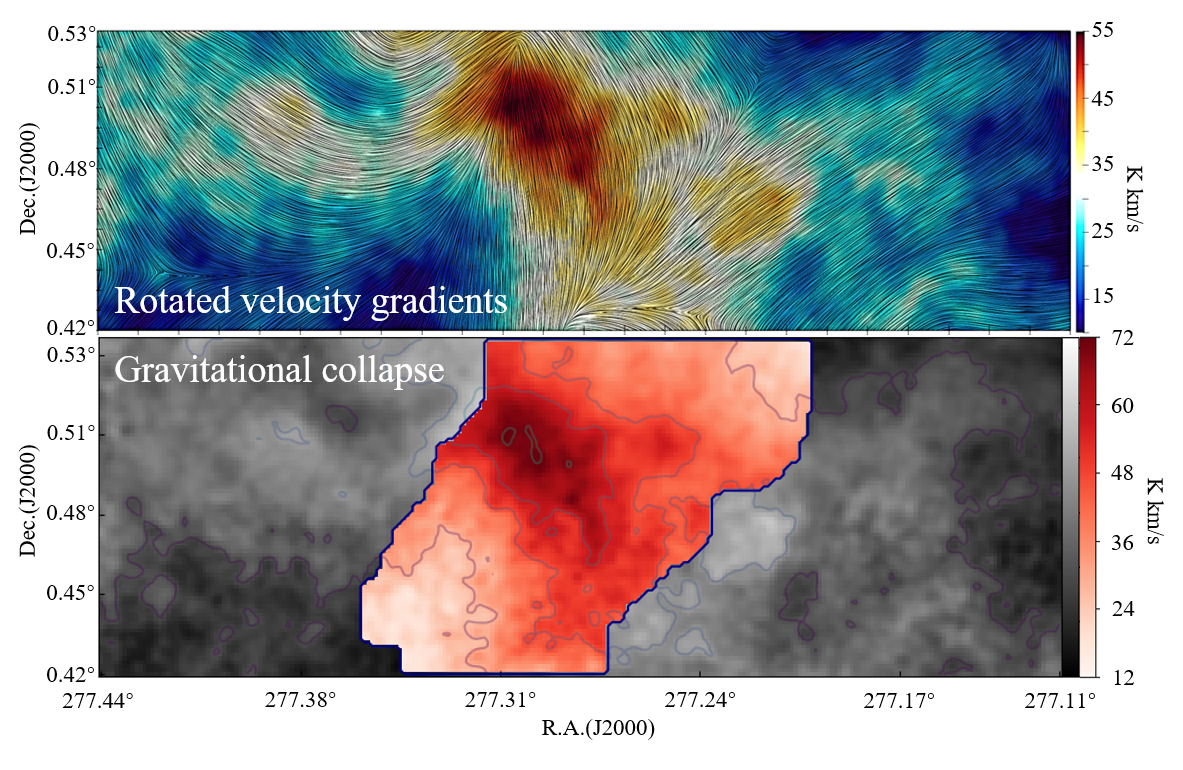}
		\caption{\label{fig:DPH} \textbf{Top:} the VGT map obtained from $\rm ^{13}CO$ (1-0) for the Serpens G3-G6 south clump. \textbf{Bottom:} the gravitational collapsing regions (red area) identified by VGT. }
	\end{figure*}
	\begin{figure*}
		\centering
		\includegraphics[width=0.825\linewidth,height=0.32\linewidth]{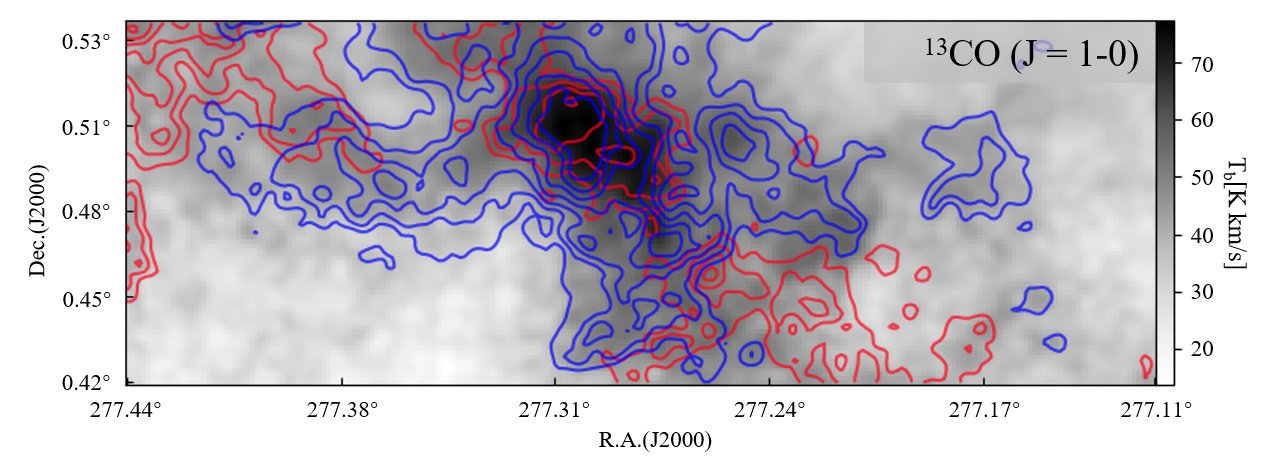}
		\caption{\label{fig:infall}Integrated intensity $\rm ^{13}CO$ (J = 1-0) emission in Serpens G3-G6 south clump. Blueshifted gas is integrated in a velocity range of +3 to +7 km $\rm s^{-1}$. Contours start at 1.5 times of the mean intensity value $\langle I_{13}^{blue}\rangle$ with a step 0.25 $\langle I_{13}^{blue}\rangle$. Redshifted gas is integrated in a velocity range of +8 to +12 km $\rm s^{-1}$. Contours start at 1.5 times of the mean intensity value $\langle I_{13}^{red}\rangle$ with a step 0.25 $\langle I_{13}^{red}\rangle$.
		}
	\end{figure*}
	
	
	\subsection{Gravitational collapsing regions identified from VGT}
	The comparison of VGT and Planck polarization directly reveals the region undergoing gravitational collapse. However, insufficient resolution in polarization measurements limits this approach in small scale studies. For instance, our high-resolution $\rm ^{13}CO$ (1-0) can measure the pixelized gradient up to $5''$, which is 120 times higher than Planck's resolution $10'$. Nevertheless, the double-peak histogram (see \S~\ref{sec:method}) extends the VGT to identify the gravitational collapsing region independent of polarization measurements.
	
	In Fig.~\ref{fig:DPH}, we present the gradients map and the identified gravitational collapsing regions using the $\rm ^{13}CO$ (1-0) emission line. In dense areas, we find the gradients flips their direction by 90$^\circ$ comparing with surrounding low-intensity regions. The DPH finds three separate gravitational collapsing regions. 
	The largest gravitational collapsing region covers the one identified by the PDFs (see Fig.~\ref{fig:NPDF}) and covers parts of low-intensity regions. It is likely the VGT is sensitive to gravity-induced inflows, which span to also low-intensity regions, while the PDFs are detecting the already formed cores. Also, the DPH cover partially diffuse regions so that the actual gravitational collapsing area is slightly overestimated. Nevertheless, both the VGT and the PDFs reveal that the central clump is gravitational collapsing. 
	
	The gravitational collapse usually shows a signature of infalling motions. Explicitly, an infall signature in a spectral line presents itself in the form of a red-blue asymmetry, generally with a diminished redshifted component \citep{1994ApJ...431..767W,1996ApJ...465L.133M}. To search for the signature, we integrated the $\rm ^{13}CO$ (J = 1-0) emission in a velocity range of +3.0 to +7.0 km $\rm s^{-1}$ for blueshifted gas, as the velocity of bulk motion is 7.65 km $\rm s^{-1}$ (see \S~\ref{sec:results}). Redshifted gas is integrated in a velocity range of +8.3 to +12.0 km $\rm s^{-1}$. In Fig.~\ref{fig:infall}, the blueshifted and redshifted gases are overlaid in integrated intensity map of $\rm ^{13}CO$ (J = 1-0). We can see the clump is dominated by blueshifted gas. In the high-intensity region, blueshifted and redshifted gases are overlapped, which indicates an infalling motion along the LOS.

	\begin{figure}
		\centering
		\includegraphics[width=1.0\linewidth,height=0.75\linewidth]{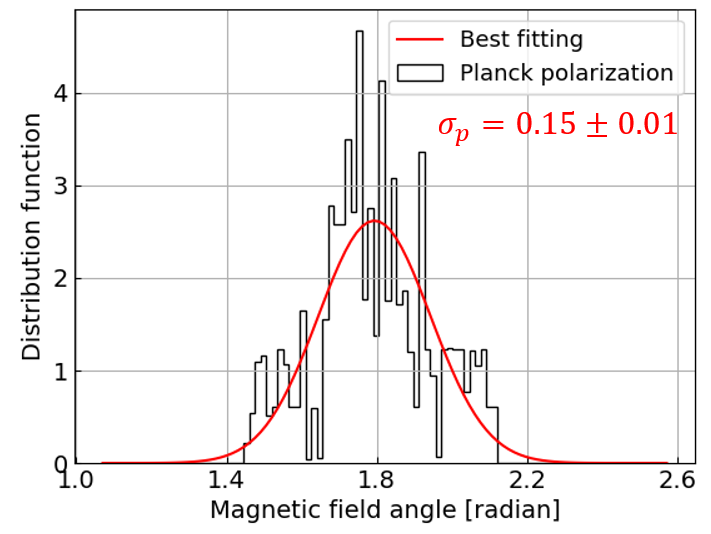}
		\caption{\label{fig:polarization} The histogram of magnetic field angle obtained from Planck 353 GHz polarization. The angle is measured in IAU convention.}
	\end{figure}
	
	\subsection{The overall energy budget of the Serpens G3-G6 south clump}
	\label{subsec:energy}
	Through both PDFs and VGT, we confirm that the Serpens G3-G6 south clump is gravitational collapsing. Here we determine the dynamics of the clump, focusing on the energy balance. The measured and derived physical parameters are listed in Tab.~\ref{tab:param}.
	\begin{table*}
		\centering
		\label{tab:param}
		\begin{tabular}{| c | c | c | c |}
			\hline
			Physical Parameter & Symbol/Definition & Value (uncertainty) & Reference\\ \hline \hline
			Area & A & 2.14 (0.20) pc$^2$ & Measured\\  
			1D velocity dispersion & $\sigma_{v,1D}$ & 1.10 (0.20) km s$^{-1}$ & Measured\\
			Polarization dispersion & $\sigma_{p}$ & 8.54 (0.72) deg& Measured\\
			H$_2$ column density & $N_0$ & 9.18 (0.43)$\times10^{21}$ $\rm cm^{-2}$ & Measured\\
			Effective diameter & $\rm L = 2 \sqrt{A/\pi}$ & 1.65 (0.08) pc & Derived\\
			Effective radius & R = L/2 & 0.83 (0.04) pc & Derived\\
			H$_2$ volume number density & $n_0=N_0/L$ & 1800.2 (117.62) $\rm cm^{-3}$ & Derived\\
			Mass of an H atom & $m_{\rm H}$ & 1.67$\times10^{-24}$ g & Ref.1 \\
			Mean molecular weight & $\mu_{\rm H_2}$ & 2.8 & Ref.1\\
			Volume mass density & $\rho_0=n_0\mu_{\rm H_2} m_{\rm H}$ & 8.42 (0.55)$\times10^{-21}$ g $\rm  cm^{-3}$ & Derived \\
			Mass & $M=N_0\mu_{\rm H_2} m_{\rm H}A$ & 440.8 (45.79) $\rm M_\odot$& Derived \\
			Magnetic field strength & $B=f\sqrt{4\pi\rho_0}\sigma_{v,1D}/\sigma_p$ & 119.88 (24.35) $\mu G$ & Derived\\
			Kinetic energy & $E_K=3M\sigma_{v,1D}^2/2$ & 1.59 $(0.60)\times10^{46}$ erg & Derived\\
			Gravitational energy & $E_G=-3GM^2/(5R)$ & -1.21 $(0.26)\times10^{46}$ erg & Derived\\
			Magnetic energy & $E_B=B^2R^3/6$ & 3.97 (1.71) $\times10^{46}$ erg & Derived\\
			Viral parameter & $\alpha_{vir}=|2E_k/E_G|$ 
			& 2.63 (1.14) & Derived\\
			Free fall time & $t_{ff}=\sqrt{3\pi/(32G\rho_0)}$ & 0.73 $(0.02)$ Myr & Derived\\
			Sound speed (isothermal)  & $c_s=\sqrt{k_BT/\mu_pm_{\rm H}}$ & 188 m s$^{-1}$& Ref.1 ($\mu_p=2.33$) \\
			Alfv\'{e}n speed & $v_A=B/\sqrt{4\pi\rho_0}$ & 3.69 (0.75) km s$^{-1}$ & Derived\\
			3D sonic Mach number & ${\rm M_S}=\sqrt{3}\sigma_{v,1D}/c_s$ & 10.1 (1.80) & Derived\\
			3D Alfv\'{e}n Mach number & ${\rm M_A}=\sqrt{3}\sigma_{v,1D}/v_A$ & 0.52 (0.14) & Derived\\
			Compressibility & $\rm \beta=2(M_A/M_S)^2$ & 0.005 (0.001) & Derived\\
			\hline
		\end{tabular}
		\caption{Physical parameters of the Serpens G3-G6 clump. All physical parameters are derived for pixels that fall within the 28.68 K km/s (mean intensity value) $\rm ^{13}CO$ (1-0) intensity contours drawn in Fig.~\ref{fig:CO_map}. All uncertainties consider the error propagation among each physical parameter. The gas temperature T is assumed to be 10 K \citep{1998ApJ...494L..19D}. The calculation of gravitational energy and magnetic energy assumes that the cloud is spherical and has a uniform density. 
			References: (1) \cite{2008A&A...487..993K}. }
	\end{table*}
	
	The area $A$ of the south clump is measured with $\rm ^{13}CO$ (1-0). We include all pixels where their integrated brightness temperature is above the mean value (see Fig.~\ref{fig:CO_map}). Adopting 415 $\pm$ 25 pc as the distance \citep{2010ApJ...718..610D}, we find $A\approx2.14$ (0.2) pc. The value in brackets indicates the uncertainty, which is the average of upper bound and lower bound. Assuming a simple spherical geometry, we have the effective LOS distance $L\approx1.65$ pc. The 1D velocity dispersion $\sigma_{v,1D}\approx1.1$ (0.2) km s$^{-1}$ is measured from the $\rm ^{13}CO$ (1-0) emission line (see Fig.~\ref{fig:CO_map}). $\sigma_{v,1D}$ contains the contribution from both turbulence velocity and shear velocity. The polarization dispersion $\sigma_p\approx8.54$ (0.72) deg is obtained from the Planck 353 GHz polarization, see Fig.~\ref{fig:polarization}. The mean H$_2$ column density $N_0\approx9.18(0.43)\times10^{21}$ $\rm cm^{-2}$.

	From these physical parameters, we derive the total mass $M=440.8$ (45.79) $\rm M_\odot$ and volume mass density $\rho_0=8.42 (0.55)\times10^{-21}$ g $\rm  cm^{-3}$. Adopting $f=0.5$, the total magnetic field strength $B$ is calculated from the Davis-Chandrasekhar-Fermi method \citep{1951PhRv...81..890D,1953ApJ...118..113C}, giving $B=f\sqrt{4\pi\rho_0}\sigma_{v,1D}/\sigma_p= 119.88 (24.35)$ $\mu G$. Assuming a spherical homogeneous cloud, the corresponding total kinetic energy $E_K= 1.59 (0.60)\times10^{46}$ erg,  total gravitational energy $E_G=-1.21 (0.26)\times10^{46}$ erg , and total magnetic field energy $E_B= 3.96 (1.71)\times10^{46}$ erg. The ratio of kinetic energy and gravitational energy $|E_K/E_G|\approx1.31$. In particular, the magnetic field energy to kinetic energy and gravitational energy ratios are $|E_B/E_K|\approx2.50$ and $|E_B/E_G|\approx3.28$, respectively. The role of magnetic field in Serpens G3-G6 south clump is much more significant than turbulence and self-gravity.
	
	Note that kinetic energy and magnetic field energy may be overestimated, as the dispersion $\sigma_{v,1D}$ considers both turbulence velocity and shear velocity. The ratio $E_B/E_K$ after simplification equivalents to  $E_B/E_K=(2f^2)/(9\sigma_p^2)$, which is independent of $\sigma_{v,1D}$. Similarly, we have $|E_B/E_G|=(10f^2 \sigma_{v,1D}^2)/(9\pi\sigma_p^2GN_0\mu_{H_2}m_HL)$. The overestimated $\sigma_{v,1D}$ leads to a stronger magnetic field and also a overestimated $|E_B/E_G|$. Nevertheless, here we have the ratio\footnote{Note in calculating total magnetic field energy, we use only the POS magnetic field component. The actual magnetic field energy can be more significant.} $|E_B/E_G|\approx3.28$, it is unlikely that the velocity dispersion is overestimated by a factor of two at least so that $|E_B/E_G|\approx1$.  Therefore, we expect the estimated magnetic field is indeed stronger than turbulence and self-gravity. The significant magnetic field energy  suggests that the gravitational collapse can happen in a strong magnetic field environment, as numerically suggested by \citet{Hu20}.\footnote{For strongly magnetized media, the reconnection diffusion, which is the consequence of turbulent reconnection (LV99), is important \citep{2005AIPC..784...42L,2012SSRv..173..557L,2010ApJ...714..442S,2020arXiv200507775S}.}  
	
	In addition, assuming the gas temperature $T \sim 10$ K \citep{1998ApJ...494L..19D}, we have the isothermal sound speed $c_s=188$ m s$^{-1}$ and Alfv\'{e}n speed $v_A=3.69 (0.75)$ km s$^{-1}$. The corresponding sonic Mach number $\rm M_S$ and Alfv\'{e}n Mach number $\rm M_A$ are therefore 10.1 (1.80) and 0.52 (0.14), respectively. Also, we have the compressibility $\beta=0.005 (0.001)$. Recall that we have the measure PDF dispersion $\sigma_s=0.55\pm0.05$ (see Fig.~\ref{fig:NPDF}). The turbulence driving parameter $b$ can be derived from Eq.~\ref{eq.b}:
	\begin{equation}
		b^2=\frac{(e^{\sigma_s^2}-1)}{\rm M_S^2}\frac{\beta+1}{\beta}
	\end{equation}
	using our derived parameters, we get $b\approx0.84$, which suggests the turbulence in Serpens G3-G6 south clump is mainly driven by compressive force. The study in \citet{2012ApJ...761..156F} shows that the supernova-driven turbulence is more effective in producing compressive motions and is expected to be more prominent in regions of enhanced stellar feedback.  Therefore, supernovas in this region may contribute to the compressive driving force here. On the other hand, solenoidal motions are more prominent in the quiescent areas with low star formation activity. Intense star formation (i.e., gravitational collapse) can also contribute, although our analysis shows that the gravity-induced motions do not dominate over large volumes. The latter point requires further studies.
	
	\subsection{Measuring mean magnetic field strength}
	The Planck measurement gives an overall view of the magnetic field. However, its resolution limits our scope to smaller scales. The dispersion of polarization measures only large scale magnetic fields, so that its value may be underestimated. The smooth field lines on a large scale suggest this underestimation is not significant. 
	
\begin{figure}
		\centering
		\includegraphics[width=1.0\linewidth,height=0.75\linewidth]{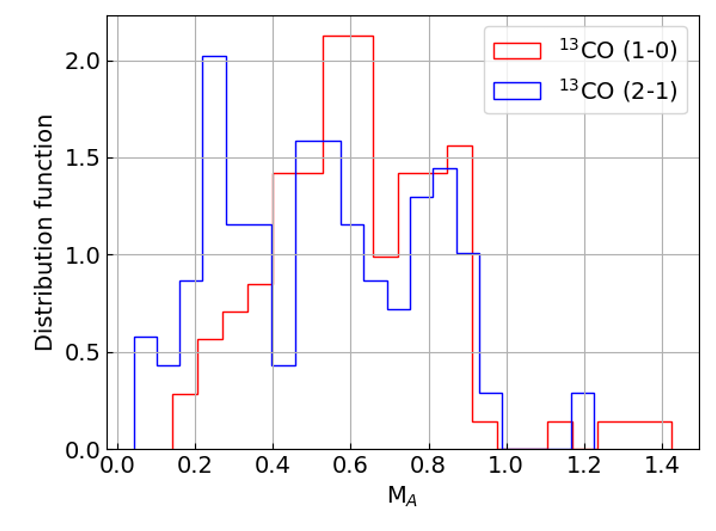}
		\caption{\label{fig:Ma}The histogram of $\rm M_A$ estimated from the distribution of velocity gradients over a sub-block \citep{2018ApJ...865...46L}. The median values of $\rm M_A$ are 0.62 and 0.53 for $\rm ^{13}CO$ (1 - 0) and $\rm ^{13}CO$ (2 - 1), respectively.}
	\end{figure}

	In addition to the DCF method, we also estimated the $\rm M_A$ directly from velocity gradients using the new approach proposed in \citet{2018ApJ...865...46L}. It was shown that the properties of velocity gradients over the sub-block is a function of $\rm M_A$. In particular, the dispersion of velocity gradients' orientation was shown to exhibit the power-law relation with $\rm M_A$ and a different power-law relation was obtained for the so-called "Top-to-Bottom" ratio of the thin channel velocity gradients (VChGs) distribution:
	\begin{equation}
		\label{eq.tb}
		\begin{aligned}
			{\rm M_A}&\approx1.6 (T_v/B_v)^{\frac{1}{-0.60\pm0.13}}, {\rm M_A}\le1\\
			{\rm M_A}&\approx7.0 (T_v/B_v)^{\frac{1}{-0.21\pm0.02}}, {\rm M_A}>1\\
		\end{aligned}
	\end{equation}
	where $T_v$ denotes the maximum value of the fitted histogram of velocity gradient's orientation, while $B_v$ is the minimum value. The analytical justification of these relations is provided in \citet{2020arXiv200207996L}. Here we, however, use the empirically obtained dependencies \citep{IGs}. This new way of obtaining Alfv\'en Mach number provides us a $\rm M_A$ distribution map for the Serpens G3-G6 south clump, as shown in Fig.~\ref{fig:Ma}.  We find the median value of $\rm M_A$ is 0.62 for the south clump. We also repeat the analysis for the $\rm ^{13}CO$ (2-1) emission line and find the median value $\rm M_A\approx0.53$. 
	
	We note that the difference of the evaluating $\rm M_A$ using the new gradient approach in \citet{2018ApJ...865...46L} and the traditional DCF method is that the new technique gets the value of $\rm M_A$ over an individual sub-block. Therefore we get not a single value of $\rm M_A$, but a distribution of magnetization of $\rm M_A$ over the cloud image. Naturally, the new way of measuring the magnetization is much more informative compared to measuring the dispersion of the projected magnetic field over the cloud image. This difference was demonstrated earlier in \citet{survey} where for a set of molecular clouds the polarization provided the mean value of magnetization and the distribution of velocity gradients provided the detailed maps of magnetization with the averaged value in good agreement with that obtained using polarization.  
	
	In the present study, we also have a similar situation. The averaged value of $\rm M_A$ estimated by the new VGT approach is close to the value that is obtained by measuring the directions of polarization. The latter is $\rm M_A=0.52\pm0.1$. This correspondence is important as the technique in \citep{2018ApJ...865...46L} and the DCF-type measurements of $\rm M_A$ are very different both in terms of the information employed and how this information is processed. This increases our confidence in our results. 
	
	Having $\rm M_A$ in hand one can use a new technique of measuring magnetic strength in \citet{2020arXiv200207996L}. The technique termed there MM2, as it uses the values of two Mach numbers, the sonic one $\rm M_S$ and the Alfv\'en one $\rm M_A$. Using the relation derived in \citet{2020arXiv200207996L} one can evaluate the POS magnetic field as:
	\begin{equation}
		B=\Omega c_s\sqrt{4\pi\rho_0}{\rm M_S M_A^{-1}},
	\end{equation}
	where $\Omega$ is a geometrical factor. By adopting $\Omega=1$ (i.e., the magnetic field perpendicular to the LOS), $\rm M_A=0.62$ (i.e., measured by VGT), $\rm M_S=10.1$, $\rho_0=8.42\times10^{-21}$g $\rm cm^{-3}$, and $c_s=188$ m $\rm s^{-1}$, we get $B\approx100\mu G$, which is close to the one ($\approx120\mu G$) derived from the DCF method.

	\section{Discussion}
	\label{sec.dis}
	\subsection{Tracing magnetic field direction with VGT}
	Measuring the magnetic field in molecular clouds is generally difficult. The polarization measurement is one practical approach to trace the magnetic fields. It is rooted in the theory of Radiative Torque alignment \citep{2007ApJ...669L..77L,Aetal15}, which predicts that the polarized dust thermal emission is perpendicular to the magnetic field and the polarized starlight is parallel to the magnetic fields. Through the measured dispersion of dust polarization's directions and the information of spectral broadening, one can estimate the magnetic field through the DCF method  \citep{2008ApJ...679..537F,2016ApJ...821...21C}. However, the polarization gives only the integrated magnetic field measurement along the LOS. 
	
	To achieve the local magnetic field measurement in molecular clouds, several techniques, for instance, the Correlation Function Analysis (CFA; \citealt{2002ASPC..276..182L,2011ApJ...740..117E}), the Structure-Function Analysis (SFA; \citealt{SFA,XH21}), and the more recent Velocity Gradients Technique (VGT; \citealt{2017ApJ...835...41G,YL17a,LY18a,PCA}), have been proposed. These techniques are based on the anisotropy of MHD turbulence, i.e., turbulent eddies are elongating along their local magnetic field direction \citep{GS95,LV99}. In particular for VGT, the velocity gradients of eddies are perpendicular to their local magnetic fields so that the gradients reveal the direction of the magnetic fields. Compared with dust polarization, the VGT, used in this work, exhibits several advantages in tracing magnetic fields. Firstly, it employs the spectroscopic data to measure the molecular cloud's local magnetic fields so that it gets rid of the contamination from the foreground \citep{LY18a,EB,2020MNRAS.496.2868L}. Using spectroscopic data and VGT, one can achieve a higher resolution of the resulting magnetic field map. Also, VGT can be applied to multiple emission lines to reveal how the magnetic field changes with the variation of density \citep{velac,PDF}. 
	
	\subsection{Identify gravitational collapsing regions by VGT}
	Observationally identifying gravitational collapsing regions is notoriously challenging. The column density PDFs provide one possible solution. In isothermal molecular clouds, the PDFs follow a log-normal distribution for non-gravitational collapsing gas. The self-gravity, however, introduces a power-law tail to the high-density range. This power-law tail statistically reveals the critical density threshold above which the gas becomes gravitational collapsing. In this work, we find the gravitational collapsing gas corresponds to density $\rm N_{H_2}\approx1.02\times10^{22}$ $\rm cm^{-2}$ in the Serpens G3-G6 clump.
	
	VGT provides an alternative way to identify the gravitational collapse. In turbulence dominated diffuse media, the velocity gradients are perpendicular to the magnetic fields \citep{LY18a}. In the presence of gravitational collapse, the gravity-induced velocity acceleration changes the velocity gradients by 90$^\circ$ being parallel to the magnetic fields \citep{YL17b,Hu20}. This change reveals the gravitational collapsing regions. For instance, VGT confirms the G3-G6 south clump is gravitational collapsing here. In addition, utilizing different emission lines, VGT can tell us the volume density range in which the collapsing occurs. As shown in Fig.~\ref{fig:HHT_vgt}, the change of gradients appears only in $\rm ^{13}CO$ map, which means the collapse happens at volume density $n\ge10^3$ $\rm cm^{-3}$.
	
	The application of VGT is not limited to nearby molecular clouds. Multiple molecular clouds and molecular filaments can be easily separated from spectroscopic PPV cubes. A complete survey of gravitational collapsing clouds and filaments is achievable using abundant spectroscopic data sets, for instance, the JCMT \citep{2019ApJ...877...43L}, GAS \citep{2017A&A...603A..89K}, COMPLETE \citep{2006AJ....131.2921R}, FCRAO \citep{1995ApJS...98..219Y}, ThrUMMS \citep{2015ApJ...812....6B}, CHaMP \citep{2005ApJ...634..476Y}, and MALT90 \citep{2011ApJS..197...25F} surveys.

	\subsection{Comparison with density gradients}
	In addition to velocity gradients, column density gradients provide an alternative way to study the magnetic fields. For instance, \citet{2013ApJ...774..128S} used the histogram of relative orientation (HRO) to characterize the relative orientation of column density gradients and the magnetic fields. The HRO technique relies on the polarimetry measurement to get magnetic field information. 
	
	The HRO should not be confused with the Intensity Gradient Technique (IGT) which is the outshoot of the VGT \citet{IGs}. The IGT can identify the direction of the magnetic field in subsonic turbulence\footnote{For some settings we know that the turbulence is subsonic. This is the case, for instance for clusters of galaxies \citep{cluster} where IGT was recently used to predict the Plane of Sky (POS) magnetic field directions.} and identify shocks in supersonic turbulence.  Different from HRO, IGT is independent of polarimetry and keeps the spatial information of density gradients by employing the sub-block averaging method, which was firstly implemented in VGT \citep{YL17a},  
	
	Here we make a comparison of HRO and IGT using the Herschel and Planck data. A numerical comparison is presented in  \citet{IGs}. For HRO, the density gradients were directly calculated for each pixel following the recipe proposed in \citet{2013ApJ...774..128S}. The gradients were then segmented according to their corresponding column density value. After that, we draw the histogram of the relative angle $\theta$ between segmented gradients and the magnetic field inferred from Planck polarization. The histogram shape parameter $\zeta = A_c - A_e$, where $A_c$ is the area under the central region of the HRO curve
	( $\frac{3}{8}\pi<\theta<\frac{5}{8}\pi$), and $A_e$ is the area in the extremes of the HRO ($0<\theta<\frac{1}{8}\pi$ and $\frac{7}{8}\pi<\theta<\pi$). As for IGT, the pixelized gradient map was further processed by the adaptive sub-block averaging method (see \S~\ref{sec:method}). A similar segmentation was also implemented to IGT and we use the AM (Eq.~\ref{AM_measure}) to quantify the performance of IGT. Note here we rotate the density gradients by 90$^\circ$ for IGT so that both positive AM and $\zeta$ indicate the {\it non-rotated} gradients are perpendicular to the magnetic field while negative AM and $\zeta$ represent a parallel relative orientation. 
	
\begin{figure}
		\centering
		\includegraphics[width=1.0\linewidth,height=0.75\linewidth]{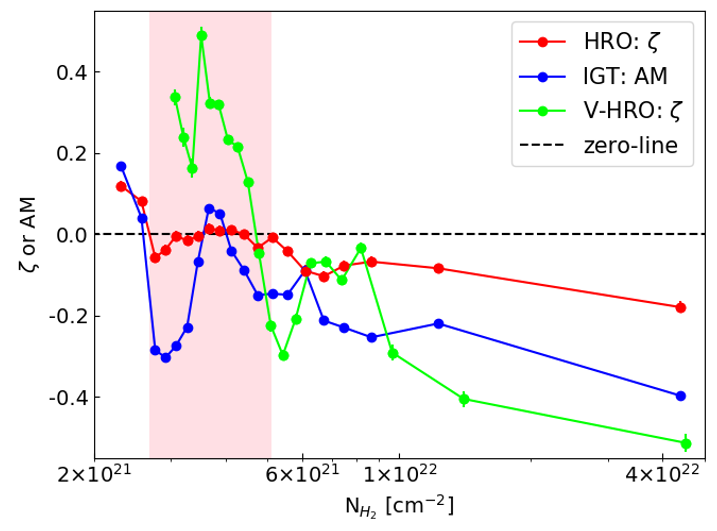}
		\caption{\label{fig:igt}The correlation between column density $\rm N_{H_2}$ and $\zeta$ or AM calculated from density gradients (red and blue) or velocity gradients (lime). Pink shadow indicate the range in which $\zeta$ of HRO is close to zero.}
	\end{figure}

	Fig.~\ref{fig:igt} presents the results of HRO and IGT for the Serpens south clump. We find both $\zeta$ and AM are positive when $\rm N_{H_2}<2.8\times10^{21}$ $\rm cm^{-2}$. $\zeta$ drops to approximately zero in the range of $\rm 2.8\times10^{21}<N_{H_2}<5.1\times10^{21}$ $\rm cm^{-2}$ and then drops to negative value further. The transition density $5.1\times10^{21}$ $\rm cm^{-2}$ is smaller than $1.01\times10^{22}$ $\rm cm^{-2}$ given by the PDFs. The behavior of AM is similar to $\zeta$, but the change is more dramatic. It is contributed by the sub-block averaging method, which enhances the statistically most important components so that density gradients can be used to trace the magnetic field in the diffuse region. The transition from positive AM and $\zeta$ to negative AM and $\zeta$ is likely caused by self-gravity, as gravitational collapse also flips the density gradients by 90$^\circ$ \citep{Hu20}. However, this change can be caused also by shocks, which make the density gradients more difficult in identifying gravitational collapsing region than velocity gradients.
	
	By analogy to the HRO for column density gradients, \citet{IGs} introduce it to the study of thin channel velocity gradients, called V-HRO. Here we use the velocity gradients calculated from $\rm ^{13}CO$ (1-0) emission line, Herschel column density data, and Planck polarization data to implement the V-HRO. The velocity gradients are calculated per pixel without the sub-block averaging method applied and are also segmented based on their corresponding column density value. We find at low-density range ($\rm N_{H_2}<4.8\times10^{21}$ $\rm cm^{-2}$), $\zeta$ of V-HRO is positive and is larger than $\zeta$ of HRO, which indicates that velocity gradients (un-rotated) have better orthogonal alignment with the magnetic fields. At the high-density range, the $\zeta$ of V-HRO drops dramatically to be negative and a flip of velocity gradients' direction occurs. As this change of direction is insensitive to shocks and it can be amplified by the sub-block averaging method, we therefore can identify the gravitational collapsing region through velocity gradients \citep{YL17b,Hu20}.

	\subsection{The interaction of turbulence, magnetic field, and self-gravity in star formation}
	Star formation serves as a hot debate in modern astrophysics. The present two classes of star-formation theory differ in the role played by magnetic fields. \citet{1956MNRAS.116..503M} firstly proposed the strong magnetic field model that clouds, which are mainly regulated by the magnetic fields, are formed with sub-critical masses. The magnetic support can be removed if mass could move across field lines. This removal is mainly conducted by neutral gas through the process of ambipolar diffusion \citep{1966MNRAS.133..265M,Spitzer1968}. The ambipolar diffusion increases the mass to flux ratio. As a consequence, cloud envelopes are subcritical, and cloud cores are supercritical. It, therefore, leads to the low efficiency of star formation. This explanation assumes flux freezing conditions of the magnetic field in the absence of turbulence. However, the turbulence is ubiquitous in ISM \citep{1995ApJ...443..209A,2010ApJ...710..853C,2017ApJ...846L..28X,2020NatAs...4.1001Z} and in the presence of turbulence the flux freezing does not hold anymore, as the process of fast turbulent magnetic reconnection breaks the magnetic freezing \citep{LV99}.  Based on the theory of turbulent reconnection,  \citet{2005AIPC..784...42L,2012SSRv..173..557L,2014SSRv..181....1L} identified reconnection diffusion (RD) as an essential process of removing magnetic flux. Consequently, the total magnetic field may be larger than its viral value, while the region can be still contracting. The reconnection diffusion (RD) theory predictions were confirmed in \citet{2010ApJ...714..442S,2012ApJ...747...21S,2013MNRAS.429.3371S} for molecular clouds and circumstellar disk settings.\footnote{The quantitative tests of the theoretical predictions related to the violation of the flux freezing and reconnection diffusion in a turbulent fluid can be found in \citet{2013Natur.497..466E} and \citet{2020arXiv200507775S}} It increases the mass to flux ratio and allows the magnetic field to equalize inside and outside the cloud, decreasing magnetic support. Unlike the ambipolar diffusion, which depends on the ionization, the reconnection diffusion is only related to the turbulent eddies' scale and the turbulent velocities \citep{2005AIPC..784...42L}. The weak magnetic field model \citep{2004RvMP...76..125M,1999ApJ...526..279P} suggests that clouds are formed at the intersection of turbulent supersonic flows. This model requires supercritical clouds so that the gravitational collapse can be triggered without removing magnetic support. 
	
	Our analysis reveals that the Serpens G3-G6 south clump is supersonic ($\rm M_S=10.1$) and sub-Alfv\'{e}nic ($\rm M_A=0.52$). The ratio $\lambda$ between the actual and critical mass-to-magnetic flux ratios $M/\Phi$ is \citep{2004mim..proc..123C}: $\lambda=7.6\times10^{-21}N_{H_2}/B\approx0.58$. $\lambda<1$ suggests the cloud is sub-critical. Also, the total magnetic field energy in the Serpens G3-G6 south clump significantly surpasses the sum of total kinetic energy and gravitational energy (see Tab.~\ref{tab:param}), suggesting the magnetic field is relatively strong in this gravitational collapsing clump. 
	
	The typical strong magnetic field model neglects the turbulence effect. Consequently, it is difficult to accumulate enough gas along the field to overcome the magnetic field \citep{1956MNRAS.116..503M}. However, turbulent compression can contribute to material accumulation. Since the strong magnetic fields provide pressures perpendicular to the field lines, the compression preferentially follows the magnetic fields and accumulates material along the perpendicular direction. Once the collapse is triggered, the inflows also predominately move along the field lines. Consequently, the gravitationally bound cloud will be thin oblate spheroids, as we see in Fig.~\ref{fig:CO_map}. Our results suggest compressive turbulence in the sub-critical Serpens G3-G6 south clump. It supports the turbulent, strong magnetic field model.
	
	Another observation of the Serpens south region's magnetic field suggests that the magnetic supercriticality sets in visual extinctions larger than 21 mag , which also indicates that gravitational collapse can be triggered in a strongly magnetized environment \citep{2020NatAs...4.1195P}.
	
	\subsection{Uncertainty and robustness}
	In this work, we use the DCF method to estimate the total magnetic field strength. The DCF method assumes that: (i) ISM turbulence is an isotropic superposition of linear small-amplitude Alfv\'{e}n waves; (ii). turbulence is homogeneous in the region studied; (iii). the compressibility and density variations of the media are negligible; (iv). the variations of the magnetic field direction and the velocity fluctuations arise from the same region in space. These assumptions could raise uncertainties in our estimation. Nevertheless, the direct measurement from Zeeman splitting reveals that the LOS magnetic field strength is $B\approx80\mu G$ for \ion{H}{1} volume density $\approx 4\times 10^3$ $\rm cm^{-3}$ \citep{2012ARAA...50..29}. As the magnetic field is amplified in gravitational collapsing region, we expect the uncertainty in our estimated POS magnetic field strength $B\approx 120\mu G$ from DCF method and $B\approx 100\mu G$ from MM2 method are not significant. 
	
	The measured velocity dispersion includes the contribution from both turbulent velocity and shear velocity. In our calculation, we did not separate the two components. As discussed in \S~\ref{subsec:energy}, the ratio between kinetic energy and magnetic energy is independent of the velocity dispersion. The uncertainty here mainly comes from polarization measurement. The velocity dispersion gives contributions to the ratio of magnetic energy and gravitational energy. However, the overestimated velocity dispersion is unlikely to compensate for the difference so that the ratio is less than one.
	
	Also, we assume the gas temperature $T\simeq10$ K. This assumption introduces uncertainties to the calculation of sound speed and sonic Mach number. The gas temperature can be roughly estimated from the optically thick $^{12}$CO (2-1) emission line \citep{2010ApJ...721..686P,2015ApJ...805...58K}: 
	\begin{equation}
		\begin{aligned}
			J_\nu(T)&=\frac{h\nu/k_B}{\exp(h\nu/(k_B T))-1}\\
			T&=\frac{h\nu_{12}}{k_B}[\log(1+\frac{h\nu_{12}/k_B}{T_{b}^{12}+J_\nu(T_{bg})})]^{-1}\\
			&=11.06[\log(1+\frac{11.06}{T_{b}^{12}+0.194})]^{-1}
		\end{aligned}
	\end{equation}
	where $h$ is the Planck constant, $\nu_{12}$ is the emission frequency of $^{12}$CO (2-1), $k_B$ is the Boltzmann constant, $T_{b}^{12}$ is the peak intensity of $^{12}$CO (2-1) in units of K, $J_\nu(T)$ is the effective radiation temperature, and $T_{bg}$ = 2.725 K is the temperature of cosmic microwave background radiation. By using a range of observed $T_b$ values, we estimate $T$ ranges from 3.7 K to 14.3 K, with a median value 7.1 K. A specie at transition level J = 1-0 is brighter than the one at J = 2-1, so it would correspond to higher temperature. We therefore, expect that the assumption $T = 10$ K does not change significantly our conclusions.
	
	\section{Conclusion} \label{sec:con}
	The energy balance between turbulence, magnetic fields, and self-gravity is still obscured in the process of star formation. In this work, we target the Serpens G3-G6 clump to study its kinetic, magnetic, and gravitational properties. We employ several data sets, including CO isotopologs’ emission lines from the HHT and IRAM telescopes, H$_2$ column density data from the Herschel Gould Belt Survey, and dust polarization data from Planck 353 GHz measurement. Our main discoveries are:
	\begin{enumerate}
		\item Using the column density PDFs and the VGT method, we confirm that the Serpens G3-G6 south clump is gravitational collapsing, which is in agreement with the observed high surface density of YSOs.
		\item The VGT method reveals that the gravitational collapse in the Serpens G3-G6 south clump occurs at volume density $n\ge10^{3}$ $\rm cm^{-3}$.
		\item We confirm that the VGT method can trace the magnetic fields and identify gravitational collapsing regions independently of polarization measurements.
		\item We use the traditional DCF method and the new MM2 technique to calculate the plane-of-the-sky magnetic field strength of the Serpens G3-G6 south clump. The magnetic field strengths estimated using the DCF method and the MM2 method are approximately 120 $\mu G$ and 100 $\mu G$, respectively.
		\item We find that the magnetic fields energy dominates the energy budget of the supersonic Serpens G3-G6 south clump, and that the turbulence is mainly driven by compressive forces.
		\item We conclude that the gravitational collapse can be successfully triggered in a supersonic and sub-Alf\'{e}nic environment. 
	\end{enumerate}
	\section*{Acknowledgements}
	A.L. acknowledges the support of the NSF grant AST 1816234 and NASA ATP AAH7546. Y.H. acknowledges the support of the NASA TCAN 144AAG1967. Flatiron Institute is supported by the Simons Foundation. This work is based on observations carried out under project number 115-19 with the IRAM 30m telescope. IRAM is supported by INSU/CNRS (France), MPG (Germany) and IGN (Spain). We acknowledge the Arizona Radio Observatory for providing the data of the Serpens region. This research has made use of data from the Herschel Gould Belt survey (HGBS) project (http://gouldbelt-herschel.cea.fr). The HGBS is a Herschel Key Programme jointly carried out by SPIRE Specialist Astronomy Group 3 (SAG 3), scientists of several institutes in the PACS Consortium (CEA Saclay, INAF-IFSI Rome and INAF-Arcetri, KU Leuven, MPIA Heidelberg), and scientists of the Herschel Science Center (HSC). We thank Ka Wai Ho and Ka Ho Yuen for helpful discussions.
	
	\software{Julia \citep{2012arXiv1209.5145B}, Paraview \citep{ayachit2015paraview}, GILDAS1/CLASS
		(\url{http://www.iram.fr/IRAMFR/GILDAS})}
	

	
	
\end{document}